\documentclass[unnumsec,webpdf,modern,large]{mam-authoring-template-arxiv}%

\usepackage{authblk}

\graphicspath{{Fig/}}

\usepackage[mathlines, switch]{lineno}
\usepackage{hyperref}
\usepackage{cleveref}
\definecolor{linkColor}{rgb}{1,0,0}
\hypersetup{pdfborder={0 0 0},colorlinks=true,urlcolor=linkColor,citecolor=linkColor}

\newcommand{\FeIBoracite}{Fe$_3$B$_7$O$_{13}$I}
\newcommand{\STO}{SrTiO$_3$}
\newcommand{\GSO}{GdScO$_3$}

\begin{document}

% Cleaned preprint version for arXiv: journal-specific metadata, history lines,
% and publisher branding have been removed while leaving article content intact.

% \journaltitle{Microscopy and Microanalysis}
% \DOI{DOI HERE}
% \copyrightyear{2026}
% \pubyear{2026}
% \access{Advance Access Publication Date: Day Month Year}
% \appnotes{Original Article}

% \firstpage{1}

%\subtitle{Subject Section}

\title[Helium-Cooled Cryogenic STEM Imaging and Ptychography]{Helium-Cooled Cryogenic STEM Imaging and Ptychography for Atomic-Scale Study of Low-Temperature Phases}
% A. Helium-Cooled Cryogenic STEM Imaging and Ptychography for Atomic-Scale Study of Low-Temperature Phases
% B. Liquid Helium-Cooled Atomic-Resolution STEM Imaging and Ptychography

\author[1,$\dagger$]{Noah Schnitzer}
\author[1,$\dagger$]{Mariana Palos}
\author[1]{Geri Topore}
\author[2]{Nishkarsh Agarwal}
\author[3]{Maya Gates}
\author[1]{Yaqi Li}
\author[2,4]{Robert Hovden}
\author[5]{Ismail El Baggari}
\author[6]{Suk Hyun Sung}
\author[1,$\ast$]{Michele Shelly Conroy}

\affil[1]{Department of Materials, London Centre for Nanotechnology, Royal School of Mines, Imperial College London, United Kingdom}
\affil[2]{Department of Materials Science and Engineering, University of Michigan, Ann
Arbor, MI 48109, United States}
\affil[3]{h-Bar Instruments, Ann Arbor, MI 48103, United States}
\affil[4]{Department of Physics, University of Michigan, Ann Arbor, MI 48109, United States}
\affil[5]{Department of Physics and Astronomy, University of British Columbia, Vancouver, BC, Canada}
\affil[6]{Michigan Center for Materials Characterization, Ann Arbor, MI 48109, United States}

\affil[ ]{\textsuperscript{$\dagger$}\,These authors contributed equally to this work.}
\affil[ ]{\textsuperscript{$\ast$}\,Corresponding author: \href{mailto:mconroy@imperial.ac.uk}{mconroy@imperial.ac.uk}}

\authormark{Schnitzer et al.}

% \received{Date}{0}{Year}
% \revised{Date}{0}{Year}
% \accepted{Date}{0}{Year}

\abstract{
Much of the exotic functionality of prime interest in quantum materials emerges from structural and electronic ground states that can only be accessed at cryogenic temperatures.
Understanding device operation therefore requires structural characterization under the same low-temperature conditions at which these functional phases exist, as room-temperature measurements often probe a different structural state. 
Achieving atomic-resolution in scanning transmission electron microscopy imaging and particularly 4D-STEM electron ptychography at liquid helium temperature has remained extremely challenging because even small amounts of drift, vibration, and thermal instability associated with the cryogen can disrupt the stringent stability requirements of atomic-resolution STEM.
In this work we demonstrate atomic-resolution STEM and  multislice electron ptychography at temperatures as low as 20 K using a commercial helium cooled holder.
We find that rapid scans and a multi-stage registration workflow are critical to reducing artifacts associated with cryogenic instability for atomic-resolution imaging, while for ptychography scan position correction including  compensation for coupling between probe aberrations and position refinement is necessary for successful reconstructions. 
Together these results establish a pathway for reliable atomic-resolution STEM and ptychography at low temperature, enabling direct visualization of structural ground states relevant to quantum technology.
}

% \boxedtext{
% \begin{itemize}
% \item Key boxed text here.
% \item Key boxed text here.
% \item Key boxed text here.
% \end{itemize}}

\maketitle

\section{Introduction}

 %--------------------------------------------------------
\begin{figure*}[!ht]
\centering 
\includegraphics[width=\textwidth,keepaspectratio]{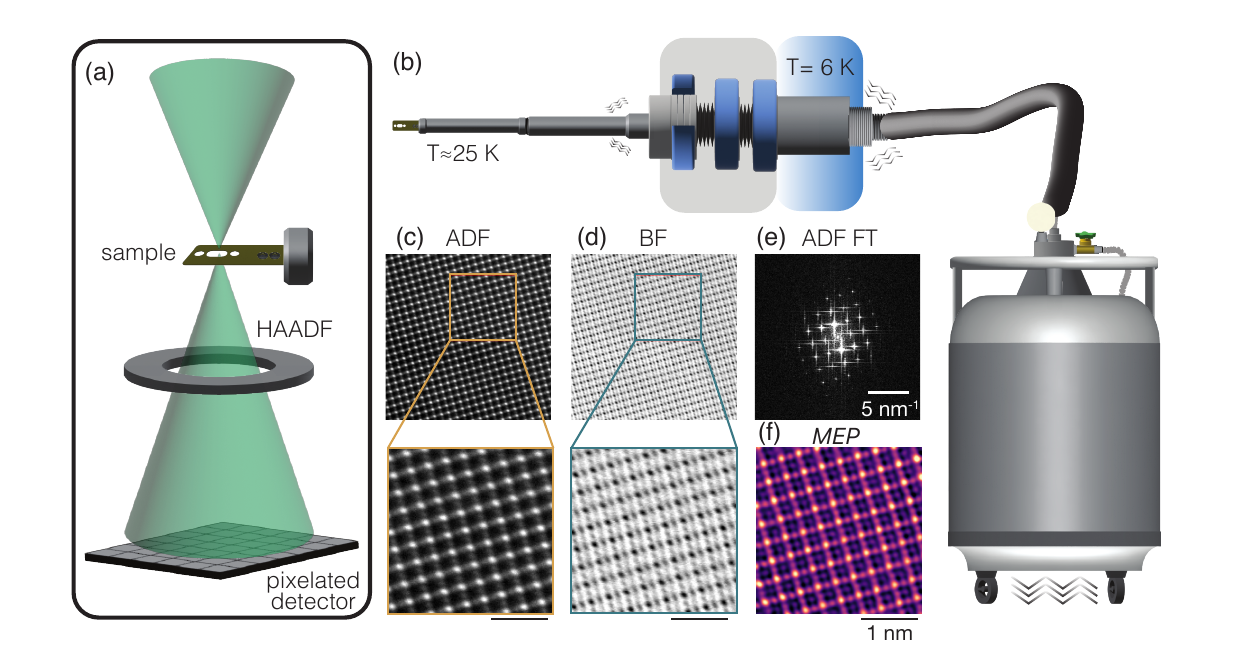}
\caption{Overview of the liquid helium atomic-resolution STEM experiment. Schematics of the (a) microscope optics and detectors and (b) sample holder, transfer line, and helium dewar illustrate the added instabilities in these conditions. The gray shaded region marks the vibration damping holder bellows, separating the holder tip from the cryostat marked with the gradient filled area. Cryostat and approximate tip temperatures for the datasets shown in (c--f) are marked. Registered and averaged images from rapid-frame (c) ADF- and (d) BF-STEM acquisitions of \FeIBoracite{} with expanded views shown below. (e) The Fourier transform of the ADF-STEM image. (f) A multislice electron ptychography reconstruction of a similar field of view to that in the expanded panels of (c) and (d), with greater sensitivity to the boron and oxygen sublattices.}
\label{Setup}
\end{figure*}
%--------------------------------------------------------

Atomic‑resolution imaging under cryogenic conditions is essential for studying structural, electronic, and ferroic phenomena that emerge only at low temperatures. Many functional oxides  including quantum paraelectrics and multiferroic materials develop symmetry lowering transitions, subtle polar distortions, and coupled order parameters when thermal energy approaches the scale of lattice and electronic instabilities\citep{ascher1966some,muller1979srti,schmid1994multi,kimura2003magnetic,Cheong/Mostovoy:2007}.
%At these low temperatures suppression of phonon activity reveals distortions and ordering phenomena that are hidden at ambient conditions. 
At these low temperatures, lattice vibrations (phonons) soften, leading to static distortions in the atomic lattice that lower the symmetry.
In ferroic oxides, the lowering of symmetry at low temperature gives rise to new domain configurations and enhanced ionic displacements that are closely linked to functional behavior. 
%In ferroic oxides, low-temperature crystallographic symmetry can differ substantially from that at room temperature, giving rise to new domain configurations and enhanced ionic displacements that are closely linked to functional behavior. 
Real-space imaging of these low-temperature phases is therefore essential both for developing materials for quantum technologies and for advancing the fundamental understanding of low-temperature structural behavior \citep{minorCryogenicElectronMicroscopy2019,collOxideElectronicsRoadmap2019,biancoAtomicResolutionCryogenicScanning2021,Cheong/Mostovoy:2007,kimura2003magnetic}.

Aberration-corrected scanning transmission electron microscopy (STEM) enables direct visualization of atomic structure and chemistry with sub-ångström spatial resolution, allowing quantitative measurement of lattice distortions and interfacial structure \citep{batsonSubangstromResolutionUsing2002,mullerAtomicScaleChemicalImaging2008}. 
The scanning nature of the technique makes minimizing vibrational coupling from the environment critical to avoid artifacts  and acquire data suitable for quantitative analysis of material structure, posing a major challenge for low temperature experiments where cryogenic cooling introduces instability \citep{mullerRoomDesignHighperformance2006,minorCryogenicElectronMicroscopy2019,biancoAtomicResolutionCryogenicScanning2021}.
Developments in fast pixelated detectors for four dimensional (4D) STEM and computational imaging methods such as electron ptychography  have provided unprecedented enhancements in spatial resolution and sensitivity to light elements. 
In ptychography, phase information encoded in overlapping diffraction discs is used to reconstruct the sample's complex transmission function from full diffraction patterns recorded at each scan position.
Because the phase retrieval relies on the redundant information of adjacent scan positions, ptychography requires exceptional mechanical and thermal stability to minimize sample drift and ensure accurate probe positioning throughout acquisition to achieve successful reconstructions \citep{maidenAnnealingAlgorithmCorrect2012,chenMixedstateElectronPtychography2020a}.

Access to low temperatures in the electron microscope is not a new interest, efforts have been made since the 1960s to use liquid nitrogen and helium cooling to access low temperature states, study phase transitions, and mitigate beam damage \citep{BLACKMAN_1963,helium1966silcox2,helium1966venerables,helium1966suzuki,HeliumreviewButler1979}.
Early foundational cryogenic transmission electron microscopy work often focused on liquid helium cooling in dedicated top down stages, \citep{Helium1963Piercy,heliumvaldre1965,helium1966silcox}, while researchers like Swann designed side-entry helium cooled holders \citep{HeliumreviewButler1979,Heliumswann70s2024}, and more recent low temperature studies have continued to develop the low temperature technique \citep{chenMicrostructureIncommensurateCommensurate1982,klie2007direct,hanHystereticResponsesSkyrmion2023,kimUltralowtemperatureCryogenicTransmission2026,mun2024atomic}.
High-resolution work has over time largely focused on liquid nitrogen cooling which is easier to work with especially in flexible side-entry systems \citep{hendersonSideentryColdHolder1991,goodgeAtomicResolutionCryoSTEMContinuously2020}.
In recent years, this has culminated in atomic-resolution STEM imaging, EELS, and electron ptychography all being successfully extended to liquid nitrogen temperatures with commercial side-entry holders \citep{biancoAtomicResolutionCryogenicScanning2021,hovdenAtomicLatticeDisorder2016,el2018nature,goodge2025direct,palos2025programmable}.
However, nitrogen cooled side-entry systems reach only $\sim$100 K, insufficient for accessing many electronic, ferroic, and spin-driven transitions of interest.

Although liquid helium cooling can reach much lower temperatures, STEM imaging, diffraction and spectroscopy in these conditions is far more limited by mechanical and thermal drift.
Previous efforts have managed to acquire atomic-resolution imaging data, but have been strongly constrained by these effects, relying on collecting data during brief stability windows \citep{klie2007direct} or restricting cryogen flow and sacrificing temperature control \citep{mun2024atomic}.

Recently, interest in side-entry liquid helium systems has accelerated with the design of a new generation of sample stages designed to minimize mechanical coupling and thermal conduction, and provide continuous helium flow control \citep{rennich2025ultracold,kimUltralowtemperatureCryogenicTransmission2026}.
With this new hardware, sub-ångström information transfer has been demonstrated in conventional TEM (CTEM) imaging \citep{sung2025liquid}, but for quantitative characterization---particularly of systems of prime interest for low temperature study characterized by subtle lattice distortions such as ferroic oxides---the flexibility and interpretability of STEM imaging modalities and emerging techniques such as electron ptychography are essential.

%In this work, we leverage these advancements in holder design to demonstrate that atomic-resolution STEM imaging and multislice electron ptychography (MEP) can be reliably performed under liquid-helium cooling conditions and stable sample temperature control below 30 K. We discuss the thermal and mechanical instabilities encountered during these measurements and how appropriate data acquisition and processing strategies can mitigate their effects. Atomic-resolution imaging is demonstrated on two low-temperature systems: a thin film of epitaxially strained \STO{} grown on (110) \GSO{}, and Fe-I boracite (\FeIBoracite{}) in the multiferroic $R3c$ phase. These examples highlight the importance of accurate registration of rapid-frame image stacks and the susceptibility of high-resolution datasets to subtle artifacts. Finally, we establish a pathway for successful multislice electron ptychography reconstruction from 4D-STEM datasets collected at liquid-helium temperatures, emphasizing the need for scan-position correction and accounting for probe aberrations. Together, these results open new opportunities to probe low-temperature states relevant to quantum technologies and to advance our understanding of low-temperature structural ground states in functional materials.

In this work, we demonstrate atomic-resolution STEM imaging and multislice electron ptychography (MEP) under stable liquid-helium cooling conditions below 30 K, using a novel cryogenic sample holder. Atomic-resolution imaging is demonstrated on two low-temperature systems: a thin film of epitaxially strained \STO{} grown on (110) \GSO{}, and Fe-I boracite (\FeIBoracite{}) in the multiferroic $R3c$ phase. 
Accurate registration of rapid-frame image stacks is key to overcoming the susceptibility of high-resolution datasets to subtle artifacts. 
We discuss the thermal and mechanical instabilities encountered during these measurements and appropriate data acquisition and processing methods to  mitigate their effects. 
Finally, we show successful multislice electron ptychography reconstruction from 4D-STEM datasets collected under liquid-helium cooling, taking advantage of the technique's self-consistent scan-position correction and identifying a coupling to probe aberrations. 
Together, these results open new opportunities to probe low-temperature states relevant to quantum technologies and to advance our understanding of low-temperature structural ground states in functional materials.

\section{Materials and Methods}
% Intro into holder (F1)
\subsection{Sample Preparation}
The cross-section lamellae of single crystal Fe-I boracite and SrTiO$_3$ grown on GdScO$_3$ were prepared for STEM imaging using a dual-beam focused ion beam integrated scanning electron microscope (Thermo-Fisher Scientific FEI G5 CX). After electron and ion beam deposition of C and Pt on the surface, the samples were thinned to electron transparency. 

\subsection{Cryogenic setup}
Samples were cooled using an h-Bar Instruments side-entry sample holder with a liquid-flow heat exchanger, supplied with liquid helium from an external dewar via a vacuum insulated transfer line.
The holder setup is shown schematically in Figure~\ref{Setup}b and detailed in \citet{rennich2025ultracold}.
Unlike conventional cryogenic sample holders in which the cryogen is contained in a small ($<$ 1~L) dewar mounted directly to the holder rod, the use of an external dewar enables long hold times and manipulation of cooling power with the helium flow rate.
On the other hand, the acoustic coupling of environmental noise to the dewar and transfer line degrades the stability of the system. 
While the design of the holder and support structure is optimized to attenuate these vibrations before the holder tip, the residual instability is the key challenge in atomic-resolution data collection in helium conditions, requiring careful data acquisition and post-processing approaches.

\subsection{Atomic-resolution STEM}
STEM datasets were collected at a 300 kV operating voltage with a convergence semi-angle of 21.4 mrad on a Thermo Fisher Scientific probe corrected XFEG Spectra 300.
Rapid frame Z-contrast annular dark field (ADF) images were collected with a 63--200 mrad collection angle range. 
4D-STEM datasets for MEP were collected with an EMPAD detector with a $\sim$1.86 ms frame time \citep{tateHighDynamicRange2016}. 
Imaging datasets were registered using the \texttt{rigidregistration} \citep{savitzkyImageRegistrationLow2018} and \texttt{quantem} \citep{ophusCorrectingNonlinearDrift2016} Python packages. 
Mixed state MEP reconstructions were performed using the \texttt{ptyrad} package \citep{chenMixedstateElectronPtychography2020a,leePtyRADHighPerformanceFlexible2025}.
Figure~\ref{Setup}c--e show simultaneous high-angle annular dark-field (HAADF) and bright-field (BF) datasets of \FeIBoracite{} acquired at a cryostat temperature of $\sim$5.5 K, corresponding to a temperature at the holder tip where the sample is mounted of approximately 20 K.
Figure~\ref{Setup}f shows the results of an MEP reconstruction of a similar region of the \FeIBoracite{} sample acquired at the same temperature.
Details on temperature measurement are provided in Supplementary Note 1.
\section{Results}

 %--------------------------------------------------------
\begin{figure*}[!ht]
\centering 
\includegraphics[width=\textwidth,keepaspectratio]{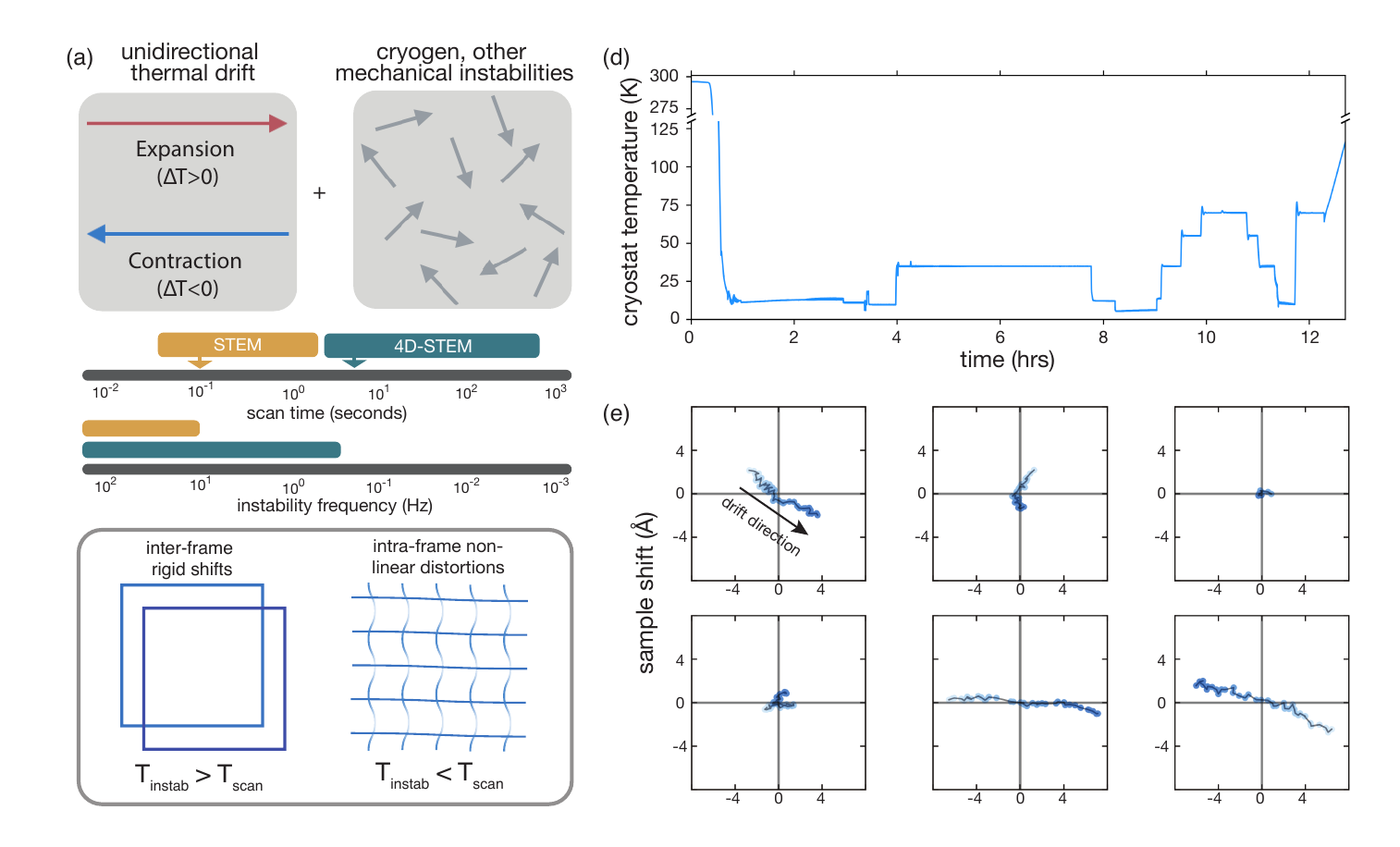}
\caption{Effects of thermal and mechanical instability on STEM imaging. 
(a) A cartoon visualization of the separation of instability in the system into large unidirectional drift associated with a temperature gradient along the holder rod (left) and the residual instabilities from the cryogen, room environment, etc. (right), not to scale. 
(b) Plot of the scan times associated with conventional  STEM imaging and 4D-STEM acquisitions, with the scan times used in this work marked with arrows. The distortion frequencies higher than the scan rate are shaded below. 
(c) Schematic illustrating the effect of instability with a frequency lower than the scan resulting in to 1st order rigid shifts between sequential frames (left), and higher frequency instability which causes distortion within individual frames, primarily on the slow scan (horizontal) direction (right). 
(d) Plot of the cryostat temperature over a day. Showing steps where the temperature was adjusted and plateaus where it was held constant for data collection.  
(e) Sample positions plotted over various 20 s time periods. Each period was optimized for atomic-resolution imaging, the cryostat temperature was constant and the stage had not been recently moved.
}
\label{instabilities}
\end{figure*}
%--------------------------------------------------------

% Instability, precise acq params (F2)
Instability during STEM imaging  will cause the electron beam to deviate from the intended scan pattern, resulting in distortions in the acquired datasets which depend strongly on the relative frequencies of the instabilities and the scan rate.
Such distortions are common even at room temperature due to electrical noise, mechanical vibration, and environmental coupling.
Under cryogenic conditions these couplings are typically amplified, and additional instability arises from cryogen flow and thermal expansion/contraction of the holder rod.
The mechanical perturbations associated with the cryogen and environmental noise primarily manifest as approximately isotropic jitter on the sub-nanometer to nanometer scale, whereas thermal expansion/contraction--- arising from either intentional change of the holder temperature or residual temperature fluctuations---produces larger-magnitude drift preferentially oriented along the holder rod (Figure~\ref{instabilities}a).

The impact of this motion on image formation depends on its characteristic frequency relative to the raster scan. 
When the instability has appreciable power at frequencies comparable to or higher than the scan line time and frame time, the resulting images exhibit complex intra-frame distortions that are most apparent along the slow-scan direction (vertical for all images in this manuscript). 
On the other hand, when the instability is slower than the frame acquisition time, individual frames can remain largely undistorted and the instability appears primarily as inter-frame translations between sequential scans (Figure~\ref{instabilities}b). 
This distinction is particularly important for quantitative analysis of atomic-resolution images, where scan artifacts can be mistaken for strains or lattice distortions.

For atomic-resolution data collection, we mitigate these distortions with a combination of hardware damping, temperature and flow control, optimization of acquisition parameters, and post-processing. 
To attenuate mechanical vibrations, the holder integrates a system of flexible bellows, and rigid supports are employed to stiffen the transfer line and support the heat exchanger (Figure~\ref{Setup}b).
To minimize thermal drift, we stabilize at a fixed temperature of interest using the stage heater and tune the helium flow to minimize temperature fluctuations at the cryostat (Fig.~\ref{instabilities}d).
Under these conditions the residual instability is dominated by remaining mechanical jitter and small temperature fluctuations.
Across multiple cooldowns we routinely achieved drift rates of 0.8--3.3 Å s$^{-1}$ during atomic-resolution imaging---comparable to liquid nitrogen imaging with side-entry holders (Fig.~\ref{instabilities}e) \citep{goodgeAtomicResolutionCryoSTEMContinuously2020}.

To recover interpretable atomic-resolution images despite significant residual instability, we rely on rapid acquisition to attempt outrun the instability.
This approach is more feasible for atomic-resolution STEM imaging on conventional integrating detectors, where we perform fast ($\sim$0.2 s) serial scans to reduce intra-frame distortion, followed by careful registration and summation to improve SNR and reduce scan artifacts.
For the 4D-STEM datasets required for ptychography, however, achievable acquisition speeds are significantly slower ($\sim$10 s) meaning the resulting scans incorporate more serious artifacts and serial acquisition is not feasible, though the irregular scan patterns can---to a limited extent---be corrected during the ptychographic reconstruction.
 %--------------------------------------------------------
\begin{figure}[!ht]
\centering 
\includegraphics[width=0.5\textwidth,height=\textheight,keepaspectratio]{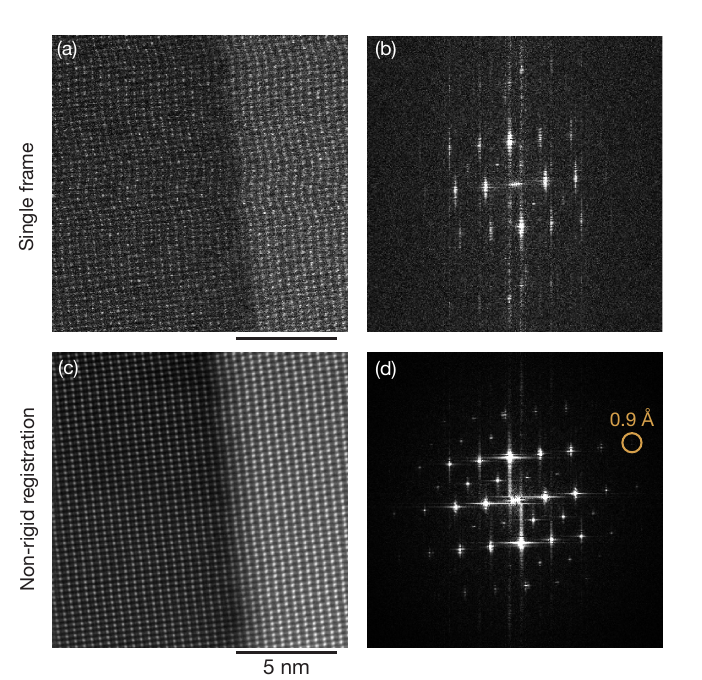}
\caption{ADF-STEM imaging performance on a \STO{} epitaxial thin film on \GSO{}. 
(a) A single rapid scan acquired at $\sim$40~K shows significant intra-frame distortion.
(b) The corresponding FFT shows significant scan artifacts, recognizable as vertical lines and anomalous peaks vertically in line with the Bragg peaks.
(c) A registered and averaged image with high signal to noise, largely free from scan distortions.
(d) The corresponding FFT, showing sub-ångström information transfer.}
\label{nrr}
\end{figure}
%--------------------------------------------------------

Figure~\ref{nrr} shows the ADF-STEM imaging performance on an epitaxial \STO{} thin film on \GSO{} acquired at a fixed temperature setpoint $\sim$40 K. 
A single rapid scan (0.14 s) across the interface (Fig.~\ref{nrr}a) clearly shows substantial frame distortion despite the short acquisition time visible in real space as sheared and wavy lattice planes.
The scan artifacts are also apparent in the corresponding fast Fourier transform (FFT) (Fig.~\ref{nrr}b), with strong vertical streaking and the appearance of noise peaks not located at reciprocal lattice points or associated with any real ordering, but characteristically aligned vertically with them.
Critically these artifacts in both real and reciprocal space are easily recognizable as they are preferentially aligned vertically along the slow scan direction, indicating that a significant component of the instability occurs on timescales comparable to the line and frame acquisition.
Rapid serial acquisition, registration, and averaging, results in a high signal image of the interface with greatly reduced distortions (Fig.~\ref{nrr}c).
The corresponding FFT (Fig.~\ref{nrr}d) shows sub-ångström information transfer, though weak residual streaking and noise peaks remain from distortions not fully corrected or averaged out through registration. 
These results demonstrate that high-resolution imaging of targeted structural features---here, the film-substrate interface---is achievable under liquid helium cooling with simultaneous control of the sample temperature, but also that the usefulness of the images depends strongly on successful series registration.

%--------------------------------------------------------
\begin{figure*}[!ht]
\centering 
\includegraphics[width=\textwidth,keepaspectratio]{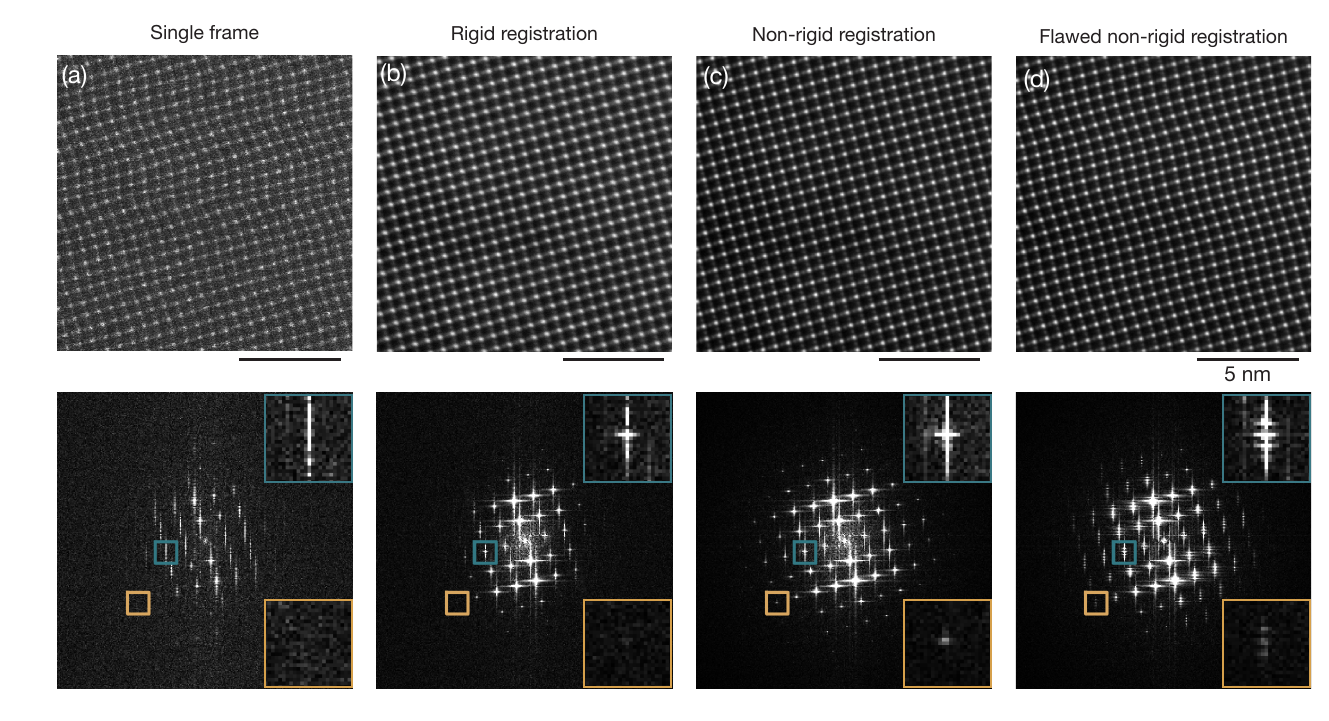}
\caption{Suitability of registration approaches applied to \FeIBoracite{}. 
(a) A single rapid scan showing severe distortions in the ADF-STEM image (top) and corresponding FFT (bottom) from instability. Insets in the FFT expand the area around two peaks: in teal, a strong reflection elongated to a line by the scan distortions, in gold a peak too faint to measure from a single scan.
(b) The result of rigid registration with optimized real and reciprocal space filtering and transitivity constrained shift refinement. Scan distortions and artifacts in the FFT are reduced, but not eliminated entirely.
(c) The result of non-rigid registration, applied to the rigidly registered result from (b). Information transfer is improved (see: gold inset) however artifacts from scan distortions present in the rigid registration result remain.
(d) The result of a flawed non-rigid registration. The image series was not correctly rigidly registered prior to non-rigid optimization. The image and FFT show similar enhanced contrast and information transfer to (c), however, very serious scan artifacts are present in the real space image and FFT.}
\label{comparisson}
\end{figure*}
%--------------------------------------------------------

Next, we present atomic-resolution imaging results on \FeIBoracite{} at $\sim$20 K in its ferroic \textit{R3c} ground-state  \citep{schnelleMagneticStructuralPhase2015}, and use this dataset to examine in detail how registration affects the quality and interpretability of atomic-resolution STEM images acquired under liquid-helium conditions. 
As in the \STO{}/\GSO{} example, instability produces severe distortions even in a single rapid scan (Fig.~\ref{comparisson}a). 
The distortion of the lattice is clear in real space, and in reciprocal space signatures of instability (streaking and artifact peaks) can be seen around a strong Bragg peak (teal inset), while due to the fast scan the low signal means the high order peaks are lost in the noise floor (gold inset).
Due to these severe intra-frame distortions, we performed rigid registration to identify the shifts between each scan, followed by non-rigid registration on the shift-corrected scans to attempt to correct distortions of each individual raster scan.

Shifts between images can be found simply by cross-correlating each image to a reference and locating the correlation maximum, whose position encodes the relative shift. 
However, for atomic-resolution imaging of crystalline materials---particularly under in situ conditions---this approach is susceptible to errors:   translational symmetry leads to ambiguous cross-correlation peaks, scans are noisy, and scan distortions introduce artifacts to the cross-correlations that can cause the algorithm to lock onto scan pathology rather than the sample structure.
To overcome these issues, we adopt the rigorous approach of \citet{savitzkyImageRegistrationLow2018} in which all scan pairs are cross-correlated and used to form a matrix of pairwise shifts that is then refined under a transitivity constraint to enforce physical consistency before being accumulated into per-frame translations, greatly reducing the incidence of mismeasured shifts. 
In this work to further mitigate the effects of the severe distortions and high noise level of our scans due to the liquid helium conditions, we adopt a two-pass Fourier filtering approach described in Supplementary Note 2 to improve the accuracy of the shift measurements. 
The resulting rigidly registered and averaged image (Fig.~\ref{comparisson}b) already suppresses most of the distortions visible in the single scan, even without any correction of intra-scan distortion. 
Key to this result is the correctness of the rigid registration---even subtly incorrect measured shifts can result in a completely unusable registered image (Supp. Fig. 3).

Following rigid registration, we apply non-rigid registration to attempt to correct the residual intra-frame distortions. 
Unlike rigid registration---where shifts are directly estimated from cross-correlations and follow a simple physical model---NRR is posed as a high-dimensional optimization problem which is harder to validate beyond post-hoc inspection for subtle artifacts, depends strongly on careful tuning of regularization parameters, and is more computationally expensive. 
When applied to the correctly rigid-registered series and properly regularized (Fig.~\ref{comparisson}c), NRR sharpens atomic columns and improves information transfer (see gold inset), but even so does not fully eliminate residual artifacts inherited from the distorted input frames. 
Critically, non-rigid registration lacks the straightforward physical consistency metrics used to identify and mitigate errors in rigid registration, making issues with non-rigid registration difficult to identify and the results difficult to validate.
Figure~\ref{comparisson}d illustrates the difficulty in identifying a failed non-rigid registration: in this case the image series was not correctly shift-corrected prior to non-rigid registration, and as a result while the non-rigid registration result appears to have converged well (sharp columns, strong contrast, apparently good information transfer), in fact the scan corrections applied are entirely incorrect. 
While in this case this is clear with close inspection---in real space the lattice planes are highly distorted alike to the single scan, and in reciprocal space strong noise peaks decorate each Bragg peak---in other cases the artifacts can be subtle, yet still must be carefully avoided to obtain images suitable for meaningful structural analysis.
Thus, while non-rigid registration can enhance image quality, particularly in the case of liquid helium STEM where intra-frame scan distortion is very difficult to avoid, great care must be taken both with the non-rigid registration itself as well as the initial rigid shift correction of scans.

 %--------------------------------------------------------
\begin{figure*}[!ht]
\centering 
\includegraphics[width=\textwidth,keepaspectratio]{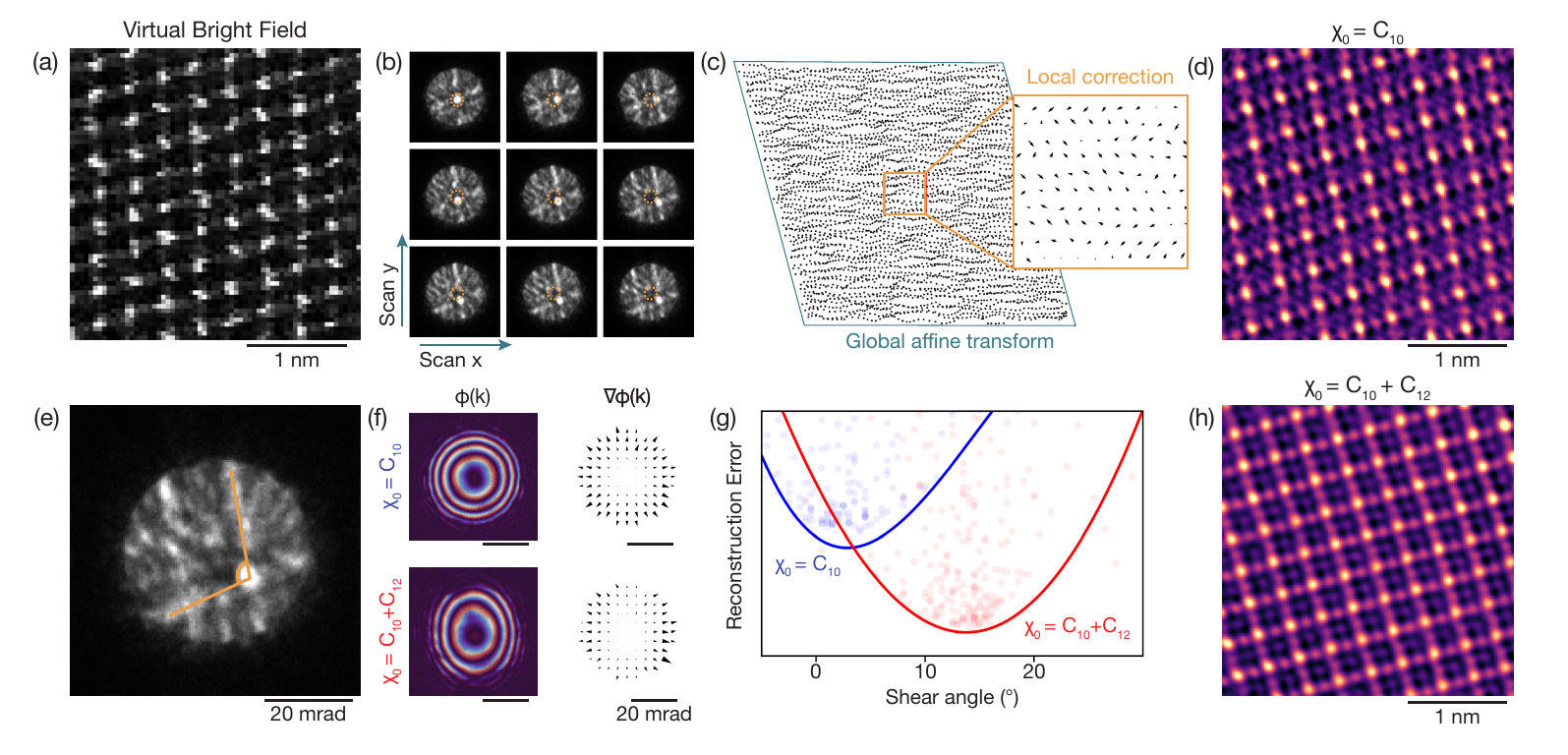}
\caption{Liquid helium multislice electron ptychography on \FeIBoracite{}. 
(a) The virtual bright field image from a defocused 64 $\times$ 64 point 4D-STEM scan. Shear in the nominally approximately square lattice of visible iodine sites is clearly visible. 
(b) A set of CBED patterns from a set of 3 $\times$ 3 points in the larger scan. An atomic column clearly visible in the shadow images formed in the bright disk of the CBED patterns moves with the scan relative to its initial position marked in orange.
(c) Corrected scan positions following a global affine transform identified from hyper-parameter tuning with Bayesian optimization, and additional gradient based individual refinement of each scan point during ptychographic reconstruction.
(d) The object phase reconstructed from a conventional probe initialized with only defocus, showing a severe shearing and distortion.
(e) A single CBED pattern annotated to show the large apparent shear angle between two nominally nearly perpendicular atomic planes.
(f) Plots of the probe phase (left) and phase gradient (right) with an aberration function including only defocus (top) and a combination of defocus and astigmatism (bottom).
(g) Reconstruction error measured for different reconstruction hyper-parameter sets, plotted as a function of shear angle applied to the scan positions. 
Results with a probe initialized with defocus only are shown in blue, while results from a probe initialized with defocus and astigmatism are plotted in red.
Lines showing the approximate optimal fronts of each setting are shown as guides for the eye.
(h) The object phase reconstructed object from a probe initialized with defocus and astigmatism, resulting in a high-quality unsheared reconstruction.
}
\label{ptycho}
\end{figure*}
%--------------------------------------------------------

Beyond conventional STEM imaging, computational imaging approaches---and in particular multislice electron ptychography (MEP)---offer significant advantages in  resolution and sensitivity compared to conventional direct imaging modes (HAADF, BF, etc.) \citep{ophusQuantitativeScanningTransmission2023}.
However, ptychography requires a 4D-STEM dataset, i.e. acquisition of a full convergent beam electron diffraction (CBED) pattern at every probe position on a pixelated detector, resulting in scans orders of magnitude slower than conventional integrating detector imaging, and posing a challenge for applying the technique in \textit{in situ} experiments.
Even acquiring with the fastest frame time available on the first-generation EMPAD \citep{tateHighDynamicRange2016} used in this work ($\sim$1.8 ms per diffraction frame, compared to pixel times on the order of $\sim 10^{-7}$~s used for rapid ADF imaging),  scans are too slow to reliably outrun distortions in liquid helium conditions.
Figure~\ref{ptycho}a shows a virtual bright-field image formed from a small $64\times64$ scan on \FeIBoracite{} in which the bright sites which should describe a nearly square lattice as in Figure~\ref{comparisson}, but instead are highly sheared from thermal drift, while the scan lines show tearing from high frequency instability.

A key advantage of ptychography, however, is that it does not require that the 4D-STEM dataset conform to a regular raster scan so long as there is sufficient overlap between sampled scan positions and the positions of the each scanned point can be identified \citep{palos2025programmable}.
By acquiring 4D-STEM datasets with a defocused probe, information about the relative positions of each scan point is encoded in the phase of the exit wave and by extension in the intensity of the recorded diffraction patterns. %, and can be used to self-consistently identify scan positions in the course of ptychographic reconstruction.
This relative positioning information can be directly seen in the motion of features in the shadow image of the central CBED disk between adjacent probe positions: in Fig.~\ref{ptycho}b a single atomic column visible in the shadow image shifts systematically with the scan relative to its initial position (orange marker).
Computationally, we exploit this redundant information by initializing the reconstruction with a nominal raster scan and then regressing probe-position updates that reduce the mismatch between the ptychographic forward model and the experimental dataset, to self-consistently identify the distorted scan positions of the 4D-STEM dataset \citep{thibaultMaximumlikelihoodRefinementCoherent2012,maidenAnnealingAlgorithmCorrect2012}. 
In practice we correct the scan positions in two stages: first, a global affine transformation is found through Bayesian optimization to account for scan-wide stretch, rotation, and shear from uniform drift during acquisition \citep{caoAutomaticParameterSelection2022}.
These transformed scan positions are used to initialize the final ptychography model, and individual scan positions are then refined during the ptychographic reconstruction to correct local distortions (Fig.~\ref{ptycho}c).
This approach can successfully compensate for a moderate degree of scan distortion, and is commonly applied for room temperature as well as recently at liquid nitrogen temperatures, but breaks down when scan-position errors are too severe to correct, e.g. when a scan position is shifted by more than a unit cell.
In this work we collected dozens of datasets, from which only a small number acquired with the smallest number of scan points trialled ($64\times64$, and thus the fastest overall scan speed, $\sim$7.3 s) successfully reconstructed.
Figure~\ref{ptycho}d shows our highest quality initial reconstruction.

Intriguingly, for liquid-helium datasets we found that these initial reconstructions exhibited an apparent shear despite an extensive parameter search for scan position correction which minimized the reconstruction error.
Further, we found that attempting to directly compensate for this shear by manipulating the scan positions degraded reconstruction quality rather than improving it, suggesting some other feature in the datasets was interfering with the position correction and preventing full convergence.
Close inspection of the shadow images in the central CBED disks showed that the lattice planes---which should nominally be nearly orthogonal---are instead significantly skewed (Fig.~\ref{ptycho}e orange overlaid marker).
As is well known (and exploited e.g. for aberration correction and tilt-corrected bright field imaging), probe aberrations distort the interference patterns of the central CBED disk, or Ronchigram, according to the gradient of the aberration phase \citep{lupiniElectronRonchigram2011,yuDoseefficientCryoelectronMicroscopy2025}.
In particular, twofold astigmatism ($C_{12}$), when combined with the defocus deliberately applied for probe overlap in ptychography, stretches the shadow image along the astigmatism axis and produces an apparent shear-like distortion (Fig.~\ref{ptycho}f).

In theory in MEP the probe should be accurately reconstructed along with the object, including astigmastism and any higher order aberrations, meaning these aberrations should have no negative effects on a fully converged reconstruction.
Practically however some degree of probe object intermixing is unavoidable due to the blind-deconvolution nature of ptychography, and decoupling probe and object features depends strongly on a good initial guess of the probe and large high signal datasets to provide high information redundancy.
Typically MEP reconstructions are initialized with a probe parametrized only by the applied defocus (i.e., neglecting asymmetric and higher order abberrations), but in our rapidly acquired datasets which are small and low signal out of necessity to minimize distortions under liquid helium conditions we find that this approach is insufficient.
Iteratively optimizing both the first order probe aberrations (C$_{10}$ and C$_{12}$) allows the astigmatism to be decoupled from the scan position correction, as shown in Figure~\ref{ptycho}g this results in a lower reconstruction error and convergence to a correct affine transform---the minimum error now corresponds to a shear angle which correctly unshears the reconstructed object. 
Our final reconstructions incorporating these corrections show the improved resolution and information transfer achievable with MEP compared to conventional imaging modes, as well as improved sensitivity to the boron and oxygen clusters key to the functional properties of the multiferroic boracite sample characterized (Fig.~\ref{ptycho}h).

\section{Discussion}
These results demonstrate that with appropriate consideration for the stability constraints associated with liquid helium cooling, atomic-resolution imaging and even multislice electron ptychography are now possible at controlled liquid helium temperatures with standard commercially available side-entry sample holders and 4D-STEM detectors.
This requires deliberate choices in data acquisition and analysis. For conventional imaging, rapid serial acquisitions are essential to minimize scan distortions, followed by robust registration approaches with physical consistency checks where possible.
For ptychography, the instability of the cryogenic condition combined with long detector frame times requires careful scan-position correction, and for small, noisy datasets additional probe aberration modeling to avoid probe–object intermixing and incorrect convergence.

At the same time, our results highlight the substantial limitations imposed by the unstable cryogenic conditions on what can be collected and reconstructed. 
Mechanical vibrations and thermal drift limit impart severe distortions even to rapid scans, limiting the datasets that can be successfully registered to small fields of view and inducing artifacts that persist even after the best achievable post-processing.
In practice, the highest-quality results were obtained only within narrow windows of optimum stability with repeated attempts, making systematic collection of datasets across varying fields of view or temperature steps difficult.

Excitingly, ongoing developments in the field promise to improve the capacity for atomic-resolution imaging at helium temperatures.
Higher speed conventional STEM imaging enabled by high-brightness sources will allow faster scans and outrunning more instability, without compromising signal or resolution \citep{goodgeFewsecondEELSMapping2021}.
Improved scan control, particularly with high-speed non-raster scan geometries may enable optimization of scan patterns for improved registration performance.
For ptychography, next-generation pixelated detectors with order-of-magnitude faster readout  will reduce frame times and improve dose efficiency per unit time, dramatically improving the quality of 4D-STEM datasets which can be collected and increasing the likelihood of successful reconstructions \citep{philippVeryHighDynamicRange2022}.
Advances in the cryogenic hardware, particularly the integration of electrical feedthroughs for control of MEMS-based devices in liquid-helium holders, promises improved temperature control without compromising the thermal stability of the system and more reliable knowledge of the sample temperature, following their successful application to liquid nitrogen holders \citep{goodgeAtomicResolutionCryoSTEMContinuously2020,schnitzer2025atomic}.
A MEMS enabled liquid helium platform would also enable more complex experiments including biasing and actuation to drive phase transitions as well as live readout of electronic properties, all in the helium-temperature regime.

On the other hand, our results also show that robust post-processing tools remain a key bottleneck. 
Most rigid and non-rigid registration algorithms are designed for high-SNR, room-temperature datasets and can fail under less ideal \textit{in situ} conditions in ways that are difficult to detect. 
Despite most registration approaches producing images that appear high quality, close examination for artifacts or of the inferred shifts reveals that the results are often dramatically incorrect, wiping out all meaningful low frequency structural information.
While for rigid registration there are approaches which when used correctly can help ensure the physicality of results, this is particularly concerning for non-rigid registration, where high-dimensional optimization and heuristic regularization do not provide intrinsic consistency checks. 
More robust, verifiable workflows are needed, ideally incorporating explicit physical constraints and uncertainty estimates to ensure that correction steps do not introduce non-physical structure.

Similar challenges apply to ptychographic scan-position correction and object-probe deconvolution in noisy \textit{in situ} datasets. While model-based position refinement is a strength of ptychography, it appears to become fragile when the dataset is small, low-SNR, or strongly distorted. 
Our observation that probe aberrations can couple to the scan geometry under liquid-helium conditions suggests a broader need for improved probe and scan position correction. 
Parametrizing probe updates with a model aberration function  \citep{yangLocalorbitalPtychographyUltrahighresolution2024a} and scan models which can represent the complex distortions encountered in \textit{in situ} datasets, such as scan line based correction, may improve robustness and the rate of successful ptychographic reconstruction in cryogenic conditions.

% Atomic-res STEM and MEP are now possible at He temperature using a commercial side-entry holder, but need to be careful with careful acquisition and reconstruction strategy: Coupled hardware + acquisition strategy + registration (not over correcting) + probe and aberration corrections

% From Fig 2: Instabilities are not random noise
%Low freq thermal drift (rod contraction/expansion)
%High freq mechanical jitter (cryo flow, acoustic coupling)
%So at He temps, intra-frame distortion is often more problematic than inter-frame drift, so multipass quick scans are ok, but more sensitive to 4DSTEM scan conditions?

%Registration:
%NRR could be visually better BUT incorrect
%Maybe treat as a refinement tool, not a primary correction

%Ptycho:

\section{Conclusions}
This work establishes a workflow and key considerations for achieving atomic-resolution STEM imaging and multislice electron ptychography at liquid-helium temperatures. 
We demonstrate high-resolution imaging with simultaneous control of sample temperature down to $\sim$20 K using a commercial side-entry holder. We characterize the instability modes that distort STEM images in these conditions, and show that rapid serial acquisition combined with carefully validated registration can recover high-SNR atomic-resolution images of targeted structural features. 
Extending beyond conventional imaging, we establish that multislice electron ptychography can be performed at liquid helium temperatures under these constraints, enabled by the self-consistent scan-position correction of ptychography and by explicitly accounting for low-order probe aberrations to avoid shear artifacts and improve convergence.

By extending high-resolution real space structural characterization to liquid-helium temperatures, these capabilities open new opportunities to directly probe low-temperature ground states across a broad range of functional and quantum materials, including strongly correlated oxides and multiferroics. 
%The ability to characterize at atomic resolution the material structure associated with exotic phenomena such as superconductivity 
Access to material structure at the low temperatures at which exotic phenomena emerge is essential both for understanding the fundamental physics of these complex phases and for guiding the design of next-generation low-power information processing and storage, and quantum technologies.

%%%%%%%%%%%%%%

\section{Competing interests}
I. E. and R. H. are inventors on a patent related to the liquid helium sample holder invention. 
I. E., R. H. and M. G. have a financial interest in h-Bar Instruments which is commercializing the invention. 
The other authors declare no competing interest.

\section{Author contributions statement}
N.S., M.P., G.T., and M.S.C. carried out STEM experiments. 
N.A., M.G., I.E., and R.H supported the cryogenic set up. 
Y.L. grew the \STO{}/\GSO{} thin film.
N.S., M.P., and G.T. analyzed the data and performed ptychographic reconstructions, with assistance from N.A. and S.H.S.
N.S, M.P, and M.S.C. wrote the manuscript, with input and review from all authors.

\section{Acknowledgments}
N.S. and M.S.C. acknowledge funding from the ERC CoG DISCO grant 101171966.  
M.P., Y.L. and M.S.C. acknowledge funding from the Royal Society Tata University Research Fellowship (URF\textbackslash R1\textbackslash 201318) and Royal Society Enhancement Award RF\textbackslash ERE\textbackslash210200EM1.
G.T. acknowledges funding from the EPSRC Centre for Doctoral Training in the Advanced Characterisation of Materials (CDTACM)(EP/S023259/1) and thank Cameca Ltd. for co-funding their PhD. 
I.E., R.H. and M.G. were supported by a National Science Foundation SBIR Phase II grant (2507716) for cryogenic hardware development.
This work was made possible with support from the Michigan Center for Materials Characterization including use of the instruments and staff assistance. The authors thank Lopa Bhatt and Chia-Hao Lee for helpful discussions on the ptychography results.

\bibliographystyle{unsrtnat}
\bibliography{reference}
\end{document}

% --- Title block ---
\begin{center}
{\Large \textbf{Supplementary Information}}\\[0.75em]
{\large Helium-Cooled Cryogenic STEM Imaging and Ptychography for Atomic-Scale Study of Low-Temperature Phases}\\[1.0em]

Noah Schnitzer$^{1,\dagger}$, Mariana Palos$^{1,\dagger}$, Geri Topore$^{1}$, Nishkarsh Agarwal$^{2}$, Maya Gates$^{3}$, Yaqi Li$^{1}$, Robert Hovden$^{2,4}$, Ismail El Baggari$^{5}$, Suk Hyun Sung$^{6}$, Michele Shelly Conroy$^{1,*}$\\[0.75em]

{\small
$^{1}$Department of Materials, London Centre for Nanotechnology, Royal School of Mines, Imperial College London, United Kingdom\\
$^{2}$Department of Materials Science and Engineering, University of Michigan, Ann Arbor, MI 48109, United States\\
$^{3}$h-Bar Instruments, Ann Arbor, MI 48103, United States\\
$^{4}$Applied Physics Program, University of Michigan, Ann Arbor, MI 48105, United States\\[0.5em]
$^{5}$Department of Physics and Astronomy, University of British Columbia, Vancouver, BC, Canada\\
$^{6}$Michigan Center for Materials Characterization, Ann Arbor, MI 48109, United States\\

$^{\dagger}$These authors contributed equally to this work.\\
$^{*}$Corresponding author: \href{mailto:mconroy@imperial.ac.uk}{mconroy@imperial.ac.uk}
}
\end{center}

%\vspace{1em}

\section*{Supplementary Note 1: Temperature Calibration}
The h-Bar instruments sample holder used in this work has two temperature sensors, one at the heat exchanger and one in the holder tip near the sample mounting position. 
During the experiments presented in this manuscript the sensor at the tip was inoperable, temperatures reported are estimates of the tip temperature based on the measured heat exchanger temperature calibrated during a separate cooldown where both sensors were operational and the temperature was stepped through several steady-state intermediate values.
The degree of deviation between the tip temperature and sample temperature is not known.

\section*{Supplementary Note 2: Rigid registration approach}
While in general ''rigid registration`` may refer to translation, rotation, and scaling transforms, here we use the term to refer specifically to identification of translational shifts between serially acquired STEM scans as the other rigid transformations are not typically relevant.
Conventionally, the shift between two images can be found directly by identifying the position of the maximum of their cross-correlation.
For atomic resolution imaging of crystalline materials, however, this approach is error-prone due to translational symmetry inherent to these images which produces nearly degenerate local maxima at each lattice point, which in the presence of scan distortions and noise can result in incorrect maxima having higher intensity than that associated with the true image shift.
Most STEM image registration algorithms align series by either taking a single scan as a reference, progressively building up a reference from the sum of the previously registered scans, or using a rolling window; all of which can perform well on high-signal to noise data with minimal scan distortions and clear low frequency features but quickly encounter issues on noisy, distorted, or highly symmetric datasets.

\begin{figure}[h]
  \centering
  \includegraphics[width=0.9\linewidth]{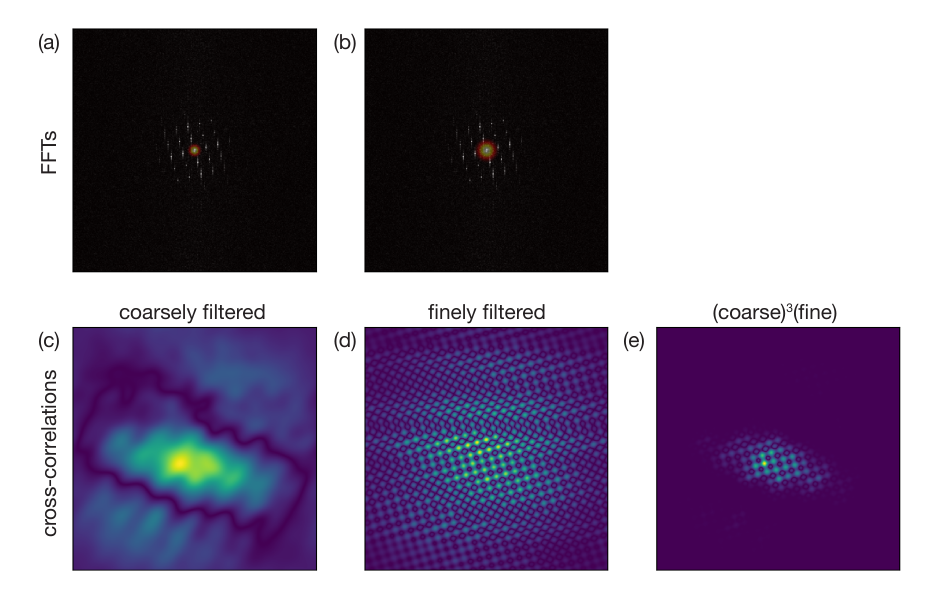}
  \caption{Coarse and fine Fourier filtering. 
  a) Fourier transform of a single scan overlaid with a ''coarse`` Fourier filter which passes only very low frequency information. 
  b) Associated coarsely filtered cross-correlation between two scans, a well defined maximum accurately marks the shift between shift between the two, but all lattice information is lost limiting precision.
  c) Fourier transform of the same scan overlaid with a ''fine`` Fourier filter which extends to the first Bragg peaks.
  d) Associated finely filtered cross-correlation, which features a local maxima at each lattice point (but is still highly filtered to minimize transfer of noise and distortions).
  e) Product of the finely filtered cross-correlation and the cube of the coarsely filtered-cross correlation, featuring a single clear maximum.
  }
  \label{fig:SF1}
\end{figure}

Our rigid registration strategy is based on that developed by \citet{savitzkyImageRegistrationLow2018}, which builds on the conventional approaches in two key ways.
Firstly, it incorporates Fourier filtering in the cross-correlation step to de-weight the high-spatial frequency information in the images, which contains only noise and highly symmetric lattice features useless for image registration, and instead add weight to the low frequencies which contain rich information like surface contamination, defects, and interfaces that contains useful information for registration.
Secondly, it sacrifices computational complexity for robustness by computing the shifts between all image pairs in the series ($(n^2-1)/2$ computations vs. $n$ to just compare each image to a reference), and uses the redundant information to check for physical consistency with transitivity and smoothness heuristics to identify and correct incorrectly measured shifts.

This approach already substantially outperforms the the alternatives, but for the highly distorted, noisy scans acquired in liquid helium conditions, further safe guards were needed to avoid incorrect measurements. 
The key parameter in the Savitzky approach is the low-pass Fourier filter used -- it must be coarse enough that noise and scan distortions are mitigated and the position of the cross-correlation maximum is accurate, but not so coarse that the lattice information is lost entirely, as ``locking-in'' to a lattice point is crucial for precision in the measured shift.
Practically, this can present an unwinnable trade-off where no single filter can be found to simultaneously provide the accuracy needed to avoid unit-cell hops from identification of the incorrect maxima without blurring out the lattice information entirely.

\begin{figure}[h]
  \centering
  \includegraphics[width=0.6\linewidth]{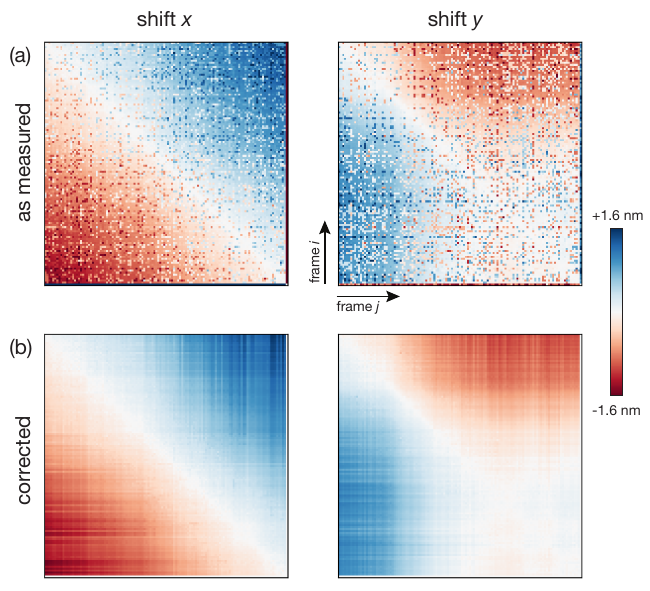}
  \caption{Pairwise shift matrices. a) \textit{x}- (left) and \textit{y}-shifts (right) measured with the two-pass all-pairs approach between each pair of images in a 100 frame series. b) Shift matrices following shift identification with transitivity and smoothness heuristics, correction with transitivity heuristic.}
  \label{fig:SF2}
\end{figure}

Here, we overcome this by performing two cross-correlation passes over all the image pairs (the ''two-pass all-pairs`` approach), one with a coarse Fourier filter which sacrifices all lattice information but entirely avoids the problem of unit-cell hops and offers highly accurate shift information exclusively from the low-frequency features of the image, and a second finer filtered pass which retains the lattice information to precisely lock-in to the nearest lattice point (Supp. Fig. \ref{fig:SF1}). 
The information from these two passes are then combined by simply multiplying the two cross-correlations -- the more coarsely filtered correlation acts as an envelope function which applies a weighting to the lattice points present in the finer correlation -- such that the true maxima has the highest intensity. 
The sharpness of the envelope function can additionally be tuned by raising this correlation to a power.
From this point the registrations follow the Savitzky approach: the shifts between all iamge pairs are assembled into shift matrices, which are refined to identify incorrect shift measurements -- unavoidable even with the careful two-pass filtering approach and visible by eye in Supp. Fig. \ref{fig:SF2}a as speckles disrupting the smooth variations of the true shifts -- and correct them  by enforcing transitivity (Supp. Fig. \ref{fig:SF2}b).

\begin{figure}[h!]
  \centering
  \includegraphics[width=0.9\linewidth]{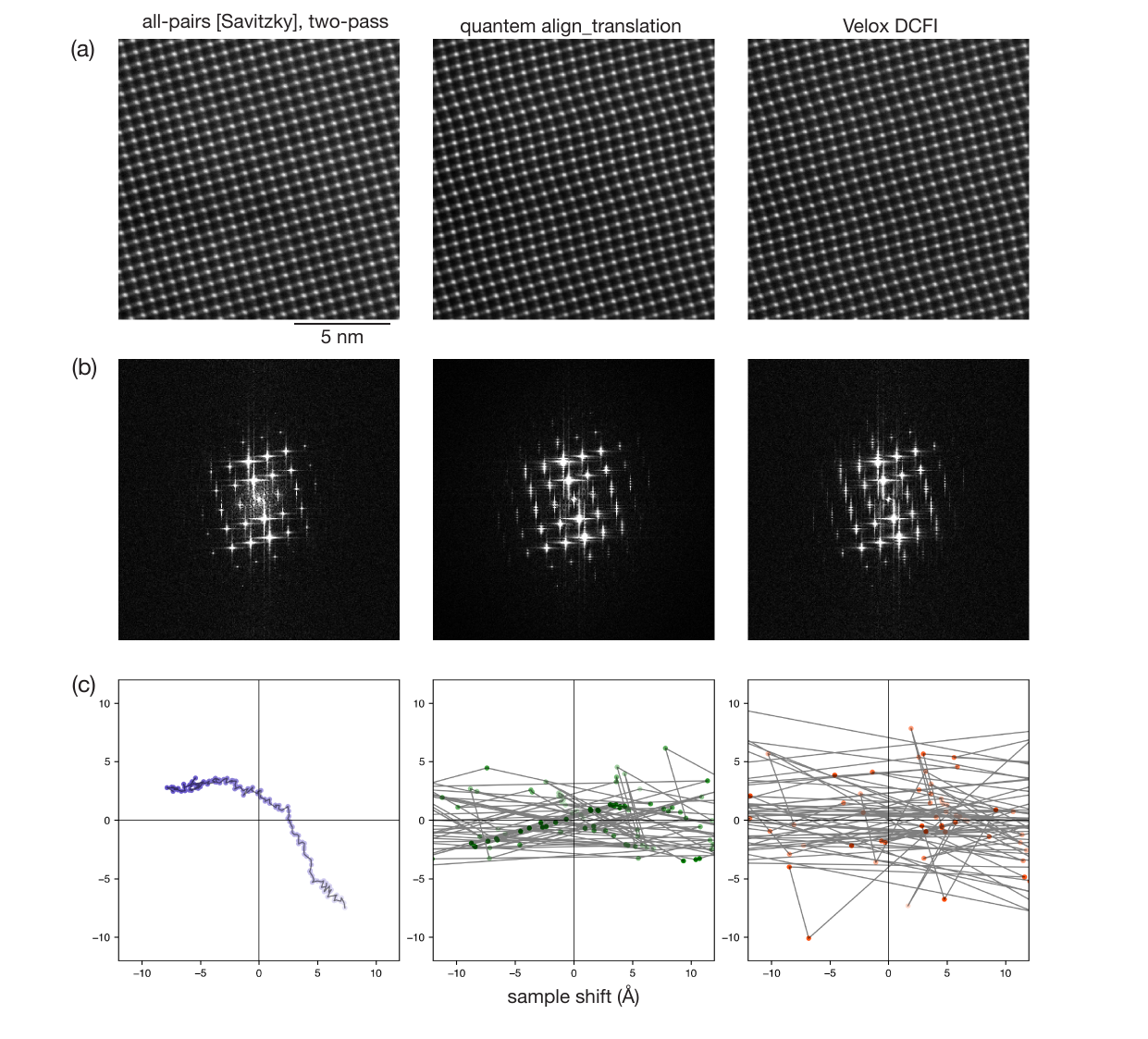}
  \caption{Comparison of rigid registration results. a) Summed images after registration with the two-pass all-pairs approach based on the Savitzky method described here (left), the \texttt{quantem} Python package's shift correction function (middle), and the ''drift-corrected frame integration`` function of the commercial Thermo Fisher Scientific Velox software operated in ''optimized for periodic images`` mode (right). b) Corresponding Fourier transforms. c) Corresponding sample drift trajectories calculated from the image shifts reported with each registration approach. A colored dot is plotted corresponding to the shift of each frame, sequential frames are connected with lines. The axes are scaled to the shifts measured with the all-pairs two-pass approach to clearly visualize the physical sample trajectory, many of the shifts measured by the other two approaches are out of frame.}
  \label{fig:SF3}
\end{figure}

The value of this approach is clear: it extracts the useful structural information present in the image series to correctly measure shifts, compared to other approaches which instead lock onto the scan distortions and noise resulting in massively incorrect results. 
This is immediately visible in comparing the artifacts present in the images and Fourier transforms resulting from other commonly applied toolkits with the Savitzky all-pairs based two-pass approach described here, and especially apparent in its clearly physical shift trajectory versus the trivially incorrect results of the other algorithms (Supp. Fig. \ref{fig:SF3}).

\bibliographystyle{unsrtnat}
\bibliography{reference} % expects references.bib